\documentclass[twocolumn,floatfix,aps]{revtex4-2}

\usepackage[T1]{fontenc}
\usepackage{amsmath}
\usepackage{amssymb}
\usepackage{amsfonts,amsthm,bm}
\usepackage{xfrac}
\usepackage[caption=false]{subfig}
\usepackage{mathrsfs}
\usepackage{tikz}
\usepackage{pgfplots}
\pgfplotsset{compat=1.18}
\usetikzlibrary{arrows.meta}
\usetikzlibrary{external}
\usepackage{standalone}
\usepackage[colorlinks=true]{hyperref}
\usepackage[sort&compress]{natbib}

\newcommand{\hamil}{\mathcal{H}}

\renewcommand{\vec}{\mathbf}

\begin{document}

\begin{abstract}
Quantum geometry has emerged as a powerful framework for understanding topological, optical and transport phenomena, revealing connections between the quantum geometric tensor (QGT) and linear response functions. Building on these developments, recent works formulate quantum geometry so that the parameter space is extended from the Brillouin zone to the space of external perturbing fields that modify the ground state. Here we develop a self-consistent diagrammatric route to determine the QGT for generic parameter spaces, including collective fluctuations or structural distortions. With this in view, we generalize previous approaches to build the geometric tensor from interacting vertex correlations that provide the spectral metric and curvature. The formalism presented here extends the interaction-dressed vertex approach (previously applied to electromagnetic responses in momentum space) to collective electronic and structural deformation fields, complementing ongoing efforts to quantify many-body contributions to quantum geometry.

\end{abstract}

%\title{Formalism to compute quantum geometrical tensor from parametrical vertices on strong-correlated systems}
%\title{Vertex-correlated formulation of many-body quantum geometry on arbitrary manifolds}\author{Alejandro S. Miñarro, Gervasi Herranz}
%\title{Generalized quantum geometry formulated through interacting vertex correlations }\author{Alejandro S. Miñarro, Gervasi Herranz}

\title{Interacting quantum geometry from dressed parametric vertices}\author{Alejandro S. Miñarro, Gervasi Herranz}

\affiliation{Institut de Ci\`encia de Materials de Barcelona (ICMAB-CSIC), Campus UAB, 08193 Bellaterra, Catalonia, Spain}
%\date{October 2025}
\maketitle

Quantum geometry provides a description of the evolution of wavefunctions as a function of parameter coordinates on a given manifold \cite{qgtguide-gao:2025, yu2025quantum, qgeomcm-liu:2025}. In solids, crystal momentum is generally taken as the parameter labeling Bloch states, defining the quantum geometric tensor (QGT) over the Brillouin zone. The real symmetric part defines the quantum metric, which quantifies how much wavefunctions differ for infinitesimal changes in momentum \cite{torma2022superconductivity, wang2023quantum, kim2025direct, kang2025measurements, sala2025quantum}. The antisymmetric imaginary part is related to the Berry curvature, which measures nontrivial geometric phases \cite{yu2025quantum, torma2023essay, qgeomcm-liu:2025}. 
The QGT can also be expressed in terms of ground-state dipole fluctuations, establishing a direct connection between quantum geometry and optical or transport response functions \cite{provost1980riemannian, you2007fidelity, ahn2022riemannian, jiang2025revealing, guan2026exploring}. This perspective endows quantum geometry with the ability to characterize the susceptibility of the ground state to external deformations \cite{verma2025quantum}. More generally, recent formulations extend the parameter space to external perturbing fields, establishing a link between the metric and response and correlation functions \cite{guan2026exploring}. This formulation therefore provides a natural route for evaluating the effects of many-body interactions in quantum geometry through the use of correlation functions and self-energy corrections \cite{sukhachov2025effect, chen-dressedberry:2022, kashihara2023quantum}. 
However, existing implementations commonly focus on momentum-space geometry and electromagnetic response channels \cite{kashihara2023quantum, chen-dressedberry:2022}. A broader diagrammatic framework is therefore needed to treat within the same quantum-geometric description interacting deformation fields, such as collective spin and orbital fluctuations or structural modes  \cite{berryjt-minarro:2026}.

\begin{figure*}[!tbh]
    \centering
    %\includestandalone[mode=build,width=0.8\linewidth]{tikz_figs/geom_distance}
    \includegraphics[width=0.8\textwidth]{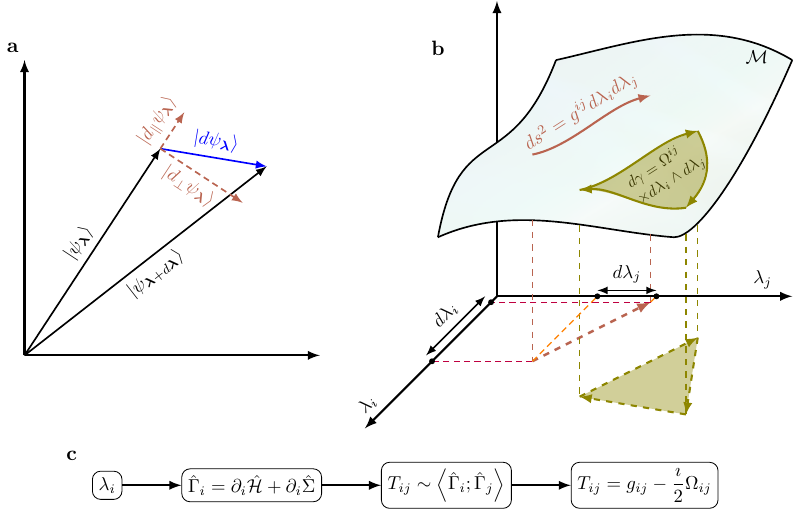}
    \caption{(a) Infinitesimal adiabatic variations of the generic parameters $\boldsymbol{\lambda}=(\lambda_1,...,\lambda_n)$ change the quantum state from $|\psi_{\boldsymbol{\lambda}}\rangle$ to $|\psi_{\boldsymbol{\lambda}+d\boldsymbol{\lambda}}\rangle$. The change $|d\psi_{\boldsymbol{\lambda}}\rangle$ can be decomposed into parallel ($|d_\parallel\psi_{\boldsymbol{\lambda}}\rangle$) and perpendicular ($|d_\perp\psi_{\boldsymbol{\lambda}}\rangle$) components with respect to the initial state $|\psi_{\boldsymbol{\lambda}}\rangle$. The perpendicular component defines the quantum geometric tensor (QGT) whose real and imaginary components are, respectively, the metric and the curvature. (b) An infinitesimal change on the parametric space, $\boldsymbol{\lambda}$ induces an infinitesimal change of wavefunctions in the manifold $\mathcal{M}$, whose distance $ds^2=||d_\perp\psi_{\boldsymbol{\lambda}}\rangle|^2$ is given by the metric $ds^2=g^{ij}d\lambda_id\lambda_j$ inherited from the pullback of the Fubini-Study metric of quantum states on the parameter manifold. On the other hand, an infinitesimal area $d\lambda_i\wedge d\lambda_j$ in the parametric space generates a geometric curvature $d\gamma=\Omega^{ij}(\boldsymbol{\lambda})d\lambda_i\wedge d\lambda_j$. The QGT $T_{ij}=g_{ij}-\frac{i}{2}\Omega_{ij}$ is defined by its metric $g_{ij}$ and curvature $\Omega_{ij}$. (c) In this work, we consider the parameter space $\vec{\lambda}$ extending to arbitrary deformation manifolds $\mathcal{M}$, where the QGT is encoded in the correlations of interacting vertices $T_{ij} \sim \left\langle \hat{\Gamma}_i ; \hat{\Gamma}_j \right\rangle$, where the vertices, defined by $\hat{\Gamma}_i=\partial_i\hat{\hamil}+\partial_i\hat{\Sigma}$, are obtained self-consistently, where $\hat{\hamil}$ is the Hamiltonian of the system, $\hat{\Sigma}$ the self-energy and the parameters $\lambda_i$ may represent, e.g., momentum, structural distortions or bosonic fluctuations.}
    \label{fig:geom-distance}
\end{figure*}

Here we develop a diagrammatic formulation of interacting quantum geometry based on the established relation between macroscopic dielectric response and the quantum geometric tensor \cite{chen-dressedberry:2022, sukhachov2025effect}. In particular, both the dressed quantum metric and Berry curvature can be obtained from the dielectric susceptibility
\(X_{ij}(\mathbf{k},i\omega_n)\), defined as the Matsubara correlator of polarization operators of the type $X_{ij}(\mathbf{k},\tau) = -\langle \hat{U}_i(\mathbf{k},\tau) \hat{U}_j^\dagger(\mathbf{k},0) \rangle$, where $\hat{U}_i(\mathbf{k}) \equiv \hat{\mathbf{c}}_{\mathbf{k}}^\dagger \hat{\mathcal{A}}_i(\mathbf{k})\hat{\mathbf{c}}_{\mathbf{k}}$ is built from non-Abelian Berry connections between $m,n$ Bloch bands $\mathcal{A}_{i,mn}(\mathbf{k})
= \langle \psi_{m\mathbf{k}}|i\partial_i|\psi_{n\mathbf{k}}\rangle$ \cite{chen-dressedberry:2022}. We have recently shown that the corresponding geometric contributions can also be extracted from the conductivity tensor $\sigma_{ij}(\vec{k},\omega) = \imath\omega\varepsilon_0 \delta_{ij} + \imath\omega X_{ij}(\vec{k},\omega)$ \cite{berryjt-minarro:2026}. Because the conductivity tensor is a response arising from analytical continuation of the correlations of current vertices through a regularized kernel, $\sigma_{ij}(\omega)\sim \langle j_i; j_j\rangle_{\omega}$ it can be viewed as a particular realization of a broader vertex-correlator structure. Here, we generalize this idea by replacing the current vertices $j_i$ with arbitrary interacting vertices $\Gamma_a$, obtaining a generalized response function $\tilde\sigma_{ab}(\omega)\sim \langle \Gamma_a;\Gamma_b\rangle_{\omega}$ (Figure \ref{fig:geom-distance}). The symmetric and antisymmetric parts of $\tilde\sigma_{ab}(\omega)$ define, respectively, the metric and curvature components of the quantum geometric tensor in a general response manifold. Thus, the conventional current vertex formulation is recovered as a particular case of the electromagnetic limit, while its general formulation allows describing other many-body phenomena incorporated through their own dressed vertex correlators, including collective (bosonic) modes or Jahn-Teller fluctuations, which are two particular cases that are briefly discussed in this work.

%A first step is to recall that the current vertex can be written in terms of the derivative with respect to momentum of the Hamiltonian and self-energy $\vec{j}_\vec{k}(t,t')\propto\boldsymbol{\nabla}_\vec{k} \left[\hat{\hamil}_\vec{k}(t)\delta(t-t') + \hat{\Sigma}_\vec{k}(t,t')  \right]$ \cite{berryjt-minarro:2026,manybody-bruus-flensberg:2004,intromanybody-coleman_chap10:2015}, i.e. the Bloch momentum vertex, its generalization is a vertex computed with the derivative of this effective Hamiltonian which includes self-energy with respect to general parameters $\boldsymbol{\lambda}$. From this generalized vertex the correlators can be also generalized in order to obtain quantum geometric tensors of any parameter space. Even, with these objects defined some general properties shared for every geometry are briefly discussed.

%beyond Bloch momentum. . e.g. the Jahn-Teller distortion space or momentum transfer, not only Bloch momentum, $\vec{k}$.

%\section{Quantum geometry fundamentals}

%\subsection{Single band quantum geometry}

We start by considering a manifold $\mathcal{M}$ of quantum states labeled by a manifold of arbitrary parameters $\boldsymbol{\lambda} = (\lambda_1,...,\lambda_n)$ (Figure \ref{fig:geom-distance}). Depending on the physical situation, $\boldsymbol{\lambda}$ may represent crystal momentum or external fields associated with strain, structural coordinates, or collective excitations that modify the state of the ground.  An infinitesimal adiabatic change $\boldsymbol{\lambda} \to \boldsymbol{\lambda} + d\boldsymbol{\lambda}$ implies a change in the quantum state $|\psi_{\boldsymbol{\lambda}+d\boldsymbol{\lambda}}\rangle = |\psi_{\boldsymbol{\lambda}}\rangle + \partial^i|\psi_{\boldsymbol{\lambda}}\rangle d\lambda_i$ \cite{qgtguide-gao:2025, yu2025quantum, qgeomcm-liu:2025}. A gauge-independent distance is defined as $ds^2 = \langle d_\perp\psi_{\boldsymbol{\lambda}}|d_\perp\psi_{\boldsymbol{\lambda}}\rangle$ where $|d_\perp\psi_{\boldsymbol{\lambda}}\rangle = \left[ 1 - |\psi_{\boldsymbol{\lambda}}\rangle\langle\psi_{\boldsymbol{\lambda}}| \right] |\partial^i \psi_{\boldsymbol{\lambda}}\rangle d\lambda_i$ is the change projected along the Hilbert space perpendicular to the initial quantum state \cite{qgtguide-gao:2025}. For systems with multiple bands, the QGT tensor is given by 
\begin{align} \label{eq:traceA}
    T_{ij}(\boldsymbol{\lambda}) &= \mathrm{tr}\left[\hat{\mathcal{A}}_i(\boldsymbol{\lambda})\hat{\mathcal{A}}_j(\boldsymbol{\lambda})\right] - \mathrm{tr}\left[\hat{\mathcal{A}}_i(\boldsymbol{\lambda})\right] \mathrm{tr}\left[\hat{\mathcal{A}}_j(\boldsymbol{\lambda})\right]
\end{align} 
where the Berry connection matrix is expressed as $\mathcal{A}_{i,\mu\nu}(\boldsymbol{\lambda}) = \imath\langle\psi_{\mu\boldsymbol{\lambda}}|\partial_i|\psi_{\nu\boldsymbol{\lambda}}\rangle$, where $\mu,\nu$ are band indices \cite{qgeomcm-liu:2025}. In the case of systems at finite temperatures, Equation \ref{eq:traceA} involves density matrices to accommodate equilibrium thermal states. The quantum metric and the Berry curvature are then given by 
\begin{subequations} \label{eq:qgt-decomp}
    \begin{align}
        g_{ij}(\boldsymbol{\lambda}) &= \Re\left[ T_{ij}(\boldsymbol{\lambda}) \right], \\
        \Omega_{ij}(\boldsymbol{\lambda}) &= -2\Im\left[ T_{ij}(\boldsymbol{\lambda}) \right].
    \end{align}
\end{subequations}

Before proceeding, let us first substantiate the geometric meaning of the parameter space when the latter is related to external deformation fields. For that, we first establish an analogy with the conventional parameter space of crystal momenta $\mathbf{k}$, where the quantum metric is the pullback of the Fubini-Study geometry to the momentum space of the Brillouin zone \cite{provost1980riemannian}. In the latter, Bloch functions $|u_n(\mathbf{k})\rangle$ are mapped by $\Phi_n:\mathrm{BZ}\longrightarrow\mathbb{P}(\mathcal{H}),\qquad \mathbf{k}\longmapsto [\,|u_n(\mathbf{k})\rangle\,]$, where $\mathbb{P}(\mathcal{H})$ is the projected Hilbert space $\mathcal{H}$. This map induces a metric in momentum space given by $g_{ij}(\mathbf{k})=\operatorname{Re}\!\left[\left\langle \partial_{k_i}u_n\left|\left(1-|u_n\rangle\langle u_n|\right)\right|\partial_{k_j}u_n\right\rangle\right]$, so the geometry in the Brillouin zone is pulled back from that of the Fubini-Study metric of the quantum states.

These concepts are extended for arbitrary external field sources, which map the parameter space $\mathcal{F}$ to the space of many-body states, i.e.
\begin{equation}
    \Phi_{0}:\mathcal{F}\rightarrow\mathbb{P}(\mathcal{H}),
    \qquad
    \vec J\mapsto [\,|\psi_0[\vec J]\rangle\,],
\end{equation}
where $\vec J=\{J_\alpha(\mathbf q)\}$ denotes the set of external fields conjugate to fermionic operators $\hat O_\alpha(-\mathbf q)$, where $\alpha$ is a component within a given interacting channel associated with, e.g., spin, charge or orbital fields. Similarly to momentum space, the geometry of the quantum states is pulled backed to the external field manifold $\mathcal{F} = \{J_\alpha(\mathbf q)\}$, endowing it with a metric, which in the zero temperature limit is given by $g_{\alpha\beta}=\operatorname{Re}T_{\alpha\beta}=\operatorname{Re}\!\left[\left\langle \partial_{J_\alpha}\psi_0\left|\left(1-|\psi_0\rangle\langle\psi_0|\right)\right|\partial_{J_\beta}\psi_0\right\rangle\right]$. Thus, the parameters $\vec J$ acquire a metric that measures the distinguishability of the states generated by infinitesimal variations of these external fields.

Interestingly, recent formulations have developed general frameworks to determine the quantum geometry of many-body systems by
treating external perturbing fields as coordinates on the space of density matrices \cite{guan2026exploring}.  Here we complement these works by outlining a self-consistent diagrammatric route that incorporates many-body self-energy and vertex corrections for general physical deformation channels, including sources coupled to collective electronic modes and structural coordinates.

To generalize these expressions, we require defining a translation generator $\delta h$ in the parameter space $\boldsymbol{\lambda} \to \boldsymbol{\lambda}+d\boldsymbol{\lambda}$ induced by a field $F_i$, $\delta h = -\imath F^i \partial_i$, so that the change in the Hamiltonian due to the translation is
\begin{align*} %\label{eq:geom-perturb}
    \delta\hat{\hamil}(\boldsymbol{\lambda}) &= \sum_{\mu,\nu} \langle \psi_{\mu\boldsymbol{\lambda}}|\delta h|\psi_{\nu\boldsymbol{\lambda}}\rangle \hat{c}^\dagger_{\mu\boldsymbol{\lambda}} \hat{c}_{\nu\boldsymbol{\lambda}} \equiv -F^i \hat{U}_i(\boldsymbol{\lambda}).
\end{align*}
Here we introduce the operator  $\hat{U}_i(\boldsymbol{\lambda}) = \hat{\vec{c}}_{\boldsymbol{\lambda}}^\dagger \hat{\mathcal{A}}_i(\boldsymbol{\lambda}) \hat{\vec{c}}_{\boldsymbol{\lambda}}$, which is related to the coupling field $\vec{F}$ through a generalized susceptibility tensor $\Upsilon_{ij}(t,t')$
\begin{subequations} \label{eq:general-susc}
    \begin{align}
        \left\langle \hat{U}_i(\boldsymbol{\lambda},t) \right\rangle &= \int_{-\infty}^t dt'\Upsilon_{ij}(\boldsymbol{\lambda};t,t') F^j(\boldsymbol{\lambda},t'), \\
        \left\langle \hat{U}_i(\boldsymbol{\lambda},\omega) \right\rangle &= \Upsilon_{ij}(\boldsymbol{\lambda};\omega) F^j(\boldsymbol{\lambda},\omega).
    \end{align}
\end{subequations}
The expression in frequency holds for equilibrium systems that are non-dispersive, i.e. $\Upsilon_{ij}(\boldsymbol{\lambda};t,t') = \Upsilon_{ij}(\boldsymbol{\lambda};t-t')$. Using the Kubo formula \cite{manybody-bruus-flensberg:2004,advancedquantum_part1-elbatanouny:2020} this susceptibility corresponds to the retarded component of a Berry connection correlator. This retarded component and the corresponding Matsubara component can be defined as
\begin{align*}
    \Upsilon_{ij}(\boldsymbol{\lambda};t,t') &= -\imath\Theta(t-t')\left\langle \left[ \hat{U}_i(\boldsymbol{\lambda},t), \hat{U}_j(\boldsymbol{\lambda},t') \right] \right\rangle \\
    \Upsilon_{ij}(\boldsymbol{\lambda},\tau) &= -\left\langle \hat{\mathcal{T}}_\tau \hat{U}_i(\boldsymbol{\lambda},\tau) \hat{U}_j(\boldsymbol{\lambda},0) \right\rangle, %= \mathrm{tr}\left[ \hat{\mathcal{A}}_i(\boldsymbol{\lambda}) \hat{G}_{\boldsymbol{\lambda}}(\tau) \hat{\mathcal{A}}_j(\boldsymbol{\lambda}) \hat{G}_{\boldsymbol{\lambda}}(-\tau) \right],
\end{align*}
where $\hat{\mathcal{T}}_\tau$ is the time-order operator for imaginary time axis. Following Ref. \cite{manybody-bruus-flensberg:2004}, we can extend the susceptibility tensor to the complex frequency plane 
\begin{align*} %\label{eq:corr-anal-cont}
    \Upsilon(z) &= -\dfrac{1}{\pi} \int_{-\infty}^\infty d\omega \dfrac{\Im[\Upsilon(\omega)]}{z-\omega},
\end{align*}
which can be continued analitically to obtain the Berry connection correlator and define a quantum geometric spectral tensor in real frequencies
\begin{align*}
    T_{ij}(\boldsymbol{\lambda},\omega) &= \dfrac{1}{2\pi}\left[ \imath\Upsilon_{ij}(\boldsymbol{\lambda},\omega) - \imath\Upsilon_{ji}^*(\boldsymbol{\lambda},\omega) \right]
\end{align*}
whose integral over all spectrum returns the proper quantum geometric tensor

\begin{align} \label{eq:spec-to-proper}
    T_{ij}(\boldsymbol{\lambda}) &= \int_0^\infty d\omega T_{ij}(\boldsymbol{\lambda},\omega).
\end{align}

The connection of Equation \ref{eq:spec-to-proper} with the geometric tensor is obtained by analogy with the correspondence between the electromagnetic response and the quantum geometry \cite{chen-dressedberry:2022}. This analogy can be naturally extended to arbitrary parameters $\boldsymbol{\lambda}$ beyond crystal momentum. 

Having established a general expression for the QGT, we now face the task of outlining a diagrammatic approach to obtain this tensor from arbitrary dressed interaction vertices $\hat{\Gamma}_i(\boldsymbol{\lambda},\tau)$. Exploiting Ward identities (to ensure gauge invariance \cite{noneqmb-stefanucci-vanLeeuwen_chap11:2013}) we parametrize this vertex as a function of $\boldsymbol{\lambda}$ as follows
\begin{align*}
    \hat{\Gamma}_i(\boldsymbol{\lambda},\tau) &= \partial_i \hat{\hamil}^{eff}_{\boldsymbol{\lambda}} = \partial_i \hat{\hamil}_{\boldsymbol{\lambda}} + \partial_i\hat{\Sigma}_{\boldsymbol{\lambda}}(\tau).
\end{align*}
where $\hat{\hamil}_{\boldsymbol{\lambda}}$ is the non-interacting Hamiltonian and $\hat{\Sigma}_{\boldsymbol{\lambda}}(\tau)$ is the self-energy \cite{manybody-bruus-flensberg:2004,quantumkin-haug-jauho:2007,electronprop-schirmer_chap4:2018}. To preserve conservation laws, the self-energy must be a functional of the Green's function, $\hat{G}_{\boldsymbol{\lambda}}(\tau)$ \cite{electronprop-schirmer_chap4:2018}, so that the vertex can be computed from a self-consistent Bethe-Salpeter equation \cite{noneqmb-stefanucci-vanLeeuwen_chap11:2013}
\begin{align} \label{eq:bethe-salpeter}
    \begin{split}
    \hat{\Gamma}_i(\boldsymbol{\lambda},\tau) &= \partial_i \hat{\hamil}_{\boldsymbol{\lambda}} + \int d\boldsymbol{\lambda}' \dfrac{\delta\hat{\Sigma}}{\delta\hat{G}_{\boldsymbol{\lambda}'}(\tau)} \\ &\phantom{=} \times \left[ \hat{G}_{\boldsymbol{\lambda}-\boldsymbol{\lambda}'}*\hat{\Gamma}_i(\boldsymbol{\lambda}-\boldsymbol{\lambda}')*\hat{G}_{\boldsymbol{\lambda}-\boldsymbol{\lambda}'} \right](\tau),
    \end{split}
\end{align}
where the symbol $*$ indicates convolution in imaginary time. Now, according to Equations (\ref{eq:general-susc}) and (\ref{eq:spec-to-proper}), we need a relation between the Berry connections and the parametric vertex, which can be obtained as follows
\begin{align}
    \imath\dfrac{\hat{U}_i(\boldsymbol{\lambda},t)}{dt} &= \left[ \hat{U}_i(\boldsymbol{\lambda},t), \hat{\hamil}_{\boldsymbol{\lambda}}^{eff}(t) \right].
\end{align}
The use of an effective Hamiltonian allows us to work with eigenfunctions $|\psi_{\mu\boldsymbol{\lambda}}\rangle$, so the matrix expansion of the Berry connection operator is $\hat{U}_i(\boldsymbol{\lambda}) = \sum_{\mu,\nu}|\psi_{\mu\boldsymbol{\lambda}}\rangle [\mathcal{A}_i(\boldsymbol{\lambda})]_{\mu\nu}\langle\psi_{\nu\boldsymbol{\lambda}}|$, so that we can express the matrix elements as follows

\begin{align} \label{eq:berry-conn-op-time-diff-mat-el}
    \begin{split}
        \langle\psi_{\mu\boldsymbol{\lambda}}| \imath\dfrac{\hat{U}_i(\boldsymbol{\lambda},t)}{dt}|\psi_{\nu\boldsymbol{\lambda}}\rangle &= [\varepsilon_{\nu\boldsymbol{\lambda}}(t) - \varepsilon_{\mu\boldsymbol{\lambda}}(t)] \\ &\phantom{=} \hspace{15pt} \times\mathcal{A}_{i,\mu\nu}(\boldsymbol{\lambda},t),
    \end{split}
\end{align}
with $\hat{\hamil}_{\boldsymbol{\lambda}}^{eff}(t)|\psi_{\mu\boldsymbol{\lambda}}\rangle = \varepsilon_{\mu\boldsymbol{\lambda}}(t)|\psi_{\mu\boldsymbol{\lambda}}\rangle$. We also need the matrix elements of the parametric vertex, which can be expressed as 
%
%the following parametric derivative, $\langle\psi_{\mu\boldsymbol{\lambda}}|\partial_i\left[ \hat{\hamil}_{\boldsymbol{\lambda}}^{eff}(t)|\psi_{\nu\boldsymbol{\lambda}}\rangle \right] = \langle\psi_{\mu\boldsymbol{\lambda}}|\partial_i\left[ \varepsilon_{\nu\boldsymbol{\lambda}}(t)|\psi_{\nu\boldsymbol{\lambda}}\rangle \right]$
%\begin{align*}    \begin{split}        \langle\psi_{\mu\boldsymbol{\lambda}}|\hat{\Gamma}_i(\boldsymbol{\lambda},t)|\psi_{\nu\boldsymbol{\lambda}}\rangle + \varepsilon_{\mu\boldsymbol{\lambda}}(t)\langle\psi_{\mu\boldsymbol{\lambda}}|\partial_i\psi_{\nu\boldsymbol{\lambda}}\rangle \\= \delta_{\mu\nu}\partial_i \varepsilon_{\nu\boldsymbol{\lambda}}(t) + \varepsilon_{\nu\boldsymbol{\lambda}}(t)\langle\psi_{\mu\boldsymbol{\lambda}}|\partial_i\psi_{\nu\boldsymbol{\lambda}}\rangle.    \end{split}\end{align*}
%Now, the vertex matrix element can be expressed as
\begin{align} \label{eq:vertex-mat-el}
    \begin{split}
        \langle\psi_{\mu\boldsymbol{\lambda}}| \hat{\Gamma}_i(\boldsymbol{\lambda},t)|&\psi_{\nu\boldsymbol{\lambda}}\rangle = \delta_{\mu\nu}\partial_i \varepsilon_{\mu\boldsymbol{\lambda}}(t) \\ &\phantom{=} - \imath[\varepsilon_{\nu\boldsymbol{\lambda}}(t)-\varepsilon_{\mu\boldsymbol{\lambda}}(t)] \mathcal{A}_{i,\mu\nu}(\boldsymbol{\lambda},t).
    \end{split}
\end{align}
Combining Equations (\ref{eq:berry-conn-op-time-diff-mat-el}) and (\ref{eq:vertex-mat-el}), and taking into account that $\partial_i \varepsilon_{\mu\boldsymbol{\lambda}}(t)$ are the diagonal elements of the parametric vertex
\begin{align} \label{eq:paramvert}
    \dfrac{d\hat{U}_i(\boldsymbol{\lambda},t)}{dt} &= \hat{\Gamma}_i(\boldsymbol{\lambda},t) - \mathrm{diag}\left[ \hat{\Gamma}_i(\boldsymbol{\lambda},t) \right]= \hat{\Gamma}^{\perp}_i(\boldsymbol{\lambda},t).
\end{align}
We therefore see that the derivative of the Berry connections $\hat{U}_i(\boldsymbol{\lambda},t)$ are the off-diagonal matrix elements of the parametric vertex defined by $\hat{\Gamma}_i^{\perp} = \hat{\Gamma}_i - \sum_n |\psi_n\rangle \langle\psi_n|\hat{\Gamma}_i|\psi_n\rangle \langle\psi_n|$. This is an important point, as $\hat{\Gamma}_i^{\perp}$ removes all non-geometric contributions to the parametric vertex in the zero temperature limit, so that $\langle\psi_m| \hat{\Gamma}_i^{\perp}|\psi_n\rangle = 0$, if $m=n$, recovering the Fubini-Study metric of quantum states.

This, in turn, allows us to obtain the time derivative of the generalized susceptibility tensor 
\begin{align}
    \begin{split}
        \dfrac{d\Upsilon_{ij}(\boldsymbol{\lambda};t,t')}{dt} &= -\imath\delta(t-t')\left\langle \left[ \hat{U}_i(\boldsymbol{\lambda},t), \hat{U}_j(\boldsymbol{\lambda},t') \right] \right\rangle \\ &\phantom{=} -\imath\Theta(t-t')\left\langle \left[ \hat{\Gamma}_i^{\perp}(\boldsymbol{\lambda},t), \hat{U}_j(\boldsymbol{\lambda},t') \right] \right\rangle.
    \end{split}
\end{align}
%
%where we have used the fact that a diagonal matrix always commutes with another matrix. This can be rewritten in frequency space as
%
%\begin{align}    \imath\omega\Upsilon_{ij}(\boldsymbol{\lambda},\omega) &= \sigma_{ij}(\boldsymbol{\lambda},\omega) + \Upsilon_{ij}(\boldsymbol{\lambda}),\end{align}
%
which can be expressed in frequency space as $\imath\omega\Upsilon_{ij}(\boldsymbol{\lambda},\omega) = \sigma_{ij}(\boldsymbol{\lambda},\omega) + \Upsilon_{ij}(\boldsymbol{\lambda})$, where we define the contact term $\Upsilon_{ij}(\boldsymbol{\lambda}) \equiv \Upsilon_{ij}(\boldsymbol{\lambda};t,t^+)$. Exploiting the cyclic property of the trace and $\mathrm{tr}[\hat{\mathcal{O}}^\dagger] = \mathrm{tr}[\hat{\mathcal{O}}]^*$, we obtain ($\hat{\rho}$ is the density matrix) 
\begin{align*}
    \begin{split}
        \Upsilon_{ij}(\boldsymbol{\lambda}) &= 2\Im\left\{ \mathrm{tr}\left[ \hat{\mathcal{A}}_i(\boldsymbol{\lambda}) \hat{\mathcal{A}}_j(\boldsymbol{\lambda})\hat{\rho}\right] \right\} \\ &=
        -2\Im\left\{ \mathrm{tr}\left[ \hat{\mathcal{A}}_j(\boldsymbol{\lambda}) \hat{\mathcal{A}}_i(\boldsymbol{\lambda})\hat{\rho}\right]\right\} = -\Upsilon_{ji}(\boldsymbol{\lambda}).
    \end{split}
\end{align*}
%
%\begin{align*}    \begin{split}         \Upsilon_{ij}(\boldsymbol{\lambda}) &= -\imath\left\langle \left[ \hat{U}_i(\boldsymbol{\lambda}), \hat{U}_j(\boldsymbol{\lambda}) \right] \right\rangle \\ &= -\imath\mathrm{tr}\left\{\left[ \hat{\mathcal{A}}_i(\boldsymbol{\lambda}) \hat{\mathcal{A}}_j(\boldsymbol{\lambda}) - \hat{\mathcal{A}}_j(\boldsymbol{\lambda}) \hat{\mathcal{A}}_i(\boldsymbol{\lambda}) \right] \hat{\rho}\right\} \\ &=        -\imath\mathrm{tr}\left[ \hat{\mathcal{A}}_i(\boldsymbol{\lambda}) \hat{\mathcal{A}}_j(\boldsymbol{\lambda})\hat{\rho} - \hat{\mathcal{A}}_i^\dagger(\boldsymbol{\lambda}) \hat{\mathcal{A}}_j^\dagger(\boldsymbol{\lambda})\hat{\rho}^\dagger \right] \\ &= 2\Im\left\{ \mathrm{tr}\left[ \hat{\mathcal{A}}_i(\boldsymbol{\lambda}) \hat{\mathcal{A}}_j(\boldsymbol{\lambda})\hat{\rho}\right] \right\} \\ &=        -2\Im\left\{ \mathrm{tr}\left[ \hat{\mathcal{A}}_j(\boldsymbol{\lambda}) \hat{\mathcal{A}}_i(\boldsymbol{\lambda})\hat{\rho}\right]\right\} = -\Upsilon_{ji}(\boldsymbol{\lambda}),    \end{split}\end{align*}
%
This implies that the QGT tensor is completely determined by the correlator $\sigma_{ij}(\boldsymbol{\lambda};t,t')=-i\Theta(t-t')\left\langle\left[\hat{\Gamma}_i(\boldsymbol{\lambda},t),\hat{U}_j(\boldsymbol{\lambda},t')\right]\right\rangle$, which in frequency space gives the spectral QGT
\begin{equation}
T_{ij}(\boldsymbol{\lambda},\omega) = \dfrac{1}{2\pi\omega}
\left[\sigma_{ij}(\boldsymbol{\lambda},\omega)+\sigma_{ji}^*(\boldsymbol{\lambda},\omega)
\right].
\label{eq:Tij-sigma}
\end{equation}
Note that the correlator $\sigma_{ij}(\boldsymbol{\lambda},\omega)$ plays a role analogous to the electrical conductivity, which allows finding the geometric curvature due to collective fluctuations \cite{berryjt-minarro:2026}. We then find the correlation between the Berry connections and the parametrized vertices as follows
\begin{align}
    \begin{split}
        \dfrac{d\sigma_{ij}(\boldsymbol{\lambda};t,t')}{dt'} &= \imath\delta(t-t')\left\langle \left[ \hat{\Gamma}_i^{\perp}(\boldsymbol{\lambda},t), \hat{U}_j(\boldsymbol{\lambda},t') \right] \right\rangle \\ &\phantom{=} -\imath\Theta(t-t')\left\langle \left[ \hat{\Gamma}_i^{\perp}(\boldsymbol{\lambda},t), \hat{\Gamma}_j^{\perp}(\boldsymbol{\lambda},t') \right] \right\rangle,
    \end{split}
\end{align}
which transformed into frequency gives $\imath\omega\sigma_{ij}(\boldsymbol{\lambda},\omega) = \chi_{ij}(\boldsymbol{\lambda},\omega) - \sigma_{ij}(\boldsymbol{\lambda})$. We introduce the perpendicular vertex correlator
\begin{equation*}
    \chi_{ij}(\boldsymbol{\lambda},\tau)
    =
    -
    \left\langle
    \hat{\mathcal{T}}_\tau
    \hat{\Gamma}_i^{\perp}(\boldsymbol{\lambda},\tau)
    \hat{\Gamma}^{\perp}_j(\boldsymbol{\lambda},0)
    \right\rangle .
\end{equation*}

which can be analytically continued according to
\begin{align}
\label{eq;anal-cont-mats}
    \begin{split}
        \chi_{ij}
        (\boldsymbol{\lambda},\imath\nu_n)
        &=
        -\frac{1}{\pi}
        \int_{-\infty}^{\infty}
        \frac{d\omega}{\imath\nu_n-\omega}
        \Im\left[
        \chi_{ij}
        (\boldsymbol{\lambda},\omega)
        \right]
        \\
        &=
        \frac{1}{\pi}
        \int_{-\infty}^{\infty}
        d\omega\,
        \frac{\omega}{\imath\nu_n-\omega}
        \Re\left[
        \sigma_{ij}(\boldsymbol{\lambda},\omega)
        \right],
    \end{split}
\end{align}
where $\imath\nu_n$ is a bosonic Matsubara frequency. The second equality holds since $\sigma_{ij}(\boldsymbol{\lambda}) = -\imath\left\langle \left[ \hat{\Gamma}_i^{\perp}(\boldsymbol{\lambda}),\hat{U}_j(\boldsymbol{\lambda}) \right] \right\rangle$ is a real tensor, which allows us to find the real part of the correlator $\Re\sigma_{ij}(\boldsymbol{\lambda},\omega)$. Using the Kramers-Kronig relations \cite{advancedquantum_part1-elbatanouny:2020}, we obtain the imaginary part of the correlator $\Im\sigma_{ij}(\boldsymbol{\lambda},\omega)$ by solving the following expression

%We introduce the vertex correlator $\chi_{ij}(\boldsymbol{\lambda},\tau) = -\left\langle \hat{\mathcal{T}}_\tau \hat{\Gamma}_i(\boldsymbol{\lambda},\tau) \hat{\Gamma}_j(\boldsymbol{\lambda},0) \right\rangle$, which can be continued analytically\begin{align} \label{eq;anal-cont-mats}    \begin{split}        \chi_{ij}(\boldsymbol{\lambda},\imath\omega_n) &= -\dfrac{1}{\pi}\int_{-\infty}^\infty \dfrac{d\omega}{\imath\omega_n-\omega}\Im\left[ \chi_{ij}(\boldsymbol{\lambda},\omega)\right] \\ &= \dfrac{1}{\pi}\int_{-\infty}^\infty d\omega \dfrac{\omega}{\imath\omega_n - \omega} \Re\left[ \sigma_{ij}(\boldsymbol{\lambda},\omega) \right].    \end{split}\end{align}

%and recalling that the QGT is given by $T_{ij}(\boldsymbol{\lambda},\omega) = \dfrac{1}{2\pi\omega}\left[ \sigma_{ij}(\boldsymbol{\lambda},\omega) + \sigma_{ji}^*(\boldsymbol{\lambda},\omega) \right]$, we finally obtain the expression of the generalized quantum geometric tensor for arbitrary manifolds and adiabatic parameter spaces, once $\sigma_{ij}(\boldsymbol{\lambda},\omega)$ is obtained by solving the following expression
\begin{align} \label{eq:kramerskronig}
    \Im\left[ \sigma_{ij}(\boldsymbol{\lambda},\omega) \right] &= -\dfrac{1}{\pi} \int_{-\infty}^\infty d\omega' \dfrac{\Re\left[ \sigma_{ij}(\boldsymbol{\lambda},\omega') \right]}{\omega'-\omega},
\end{align}

%\section{Computation of parametric vertex correlator in imaginary time axis}

\begin{figure*}[!tbh]
    \centering
    %\includestandalone[mode=build,width=0.8\textwidth]{tikz_figs/bubble_diagram}
    \includegraphics[width=0.8\textwidth]{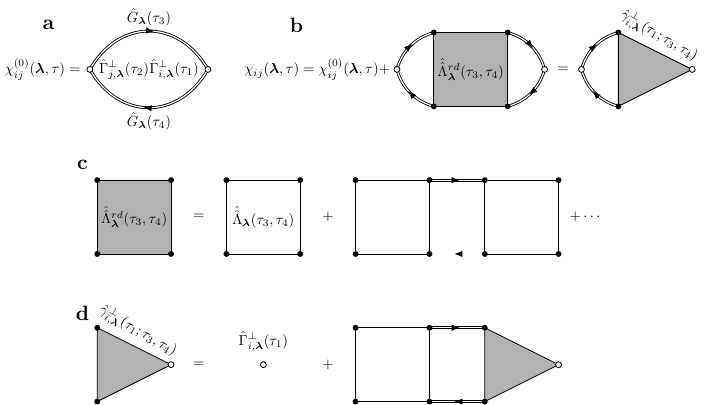}
    \caption{Diagrammatic expansion of the parametric vertex correlation bubble diagram $\chi_{ij}(\boldsymbol{\lambda},\tau)$. In panel (a) we represent the zeroth order expansion corresponding to the mean-field approximation which does not take into account the bubble interactions. These bubble (or particle-hole) interactions are introduced in panel (b) with the reducible diagram $\hat{\hat{\Lambda}}^{rd}_{\boldsymbol{\lambda}}(\tau_3,\tau_4)$. This diagram can be simplified to irreducible diagram ($\hat{\hat{\Lambda}}_{\boldsymbol{\lambda}}(\tau_3,\tau_4)$) expansion in panel (c). After the expansion, the bubble diagram is written compactly with the introduction of the three point vertex $\hat{\gamma}_i^\perp(\boldsymbol{\lambda},\tau_1;\tau_3,\tau_4)$, whose self-consistent solution (Equation \ref{eq:3p-vertex-dyson}) is represented diagrammatically in panel (d).}
    \label{fig:bubble-diagram-expansion}
\end{figure*}

Note that the crucial element to obtain the generalized quantum geometric tensor is the perpendicular parametric vertex correlator $\chi_{ij}(\boldsymbol{\lambda},\tau) = -\left\langle \hat{\mathcal{T}}_\tau \hat{\Gamma}_i^{\perp}(\boldsymbol{\lambda},\tau)\hat{\Gamma}_j^{\perp}(\boldsymbol{\lambda},0) \right\rangle$ shown in Equation \ref{eq:vertex-correl-expand}, which can be expanded using two particle Green's functions \cite{noneqmb-stefanucci-vanLeeuwen_chap5:2013} 
\begin{align} \label{eq:vertex-correl-expand}
    \begin{split}
        \chi_{ij}(\boldsymbol{\lambda},\tau) &= - \iint d\tau' d\tau''\\ &\phantom{=} \times \Gamma_{i,\mu\nu}^{\perp}(\boldsymbol{\lambda},\tau-\tau')\Gamma_{j,\mu'\nu'}^{\perp}(\boldsymbol{\lambda},-\tau'') \\ &\phantom{=}
        \times\left\langle \hat{\mathcal{T}}_\tau \hat{c}_{\mu\boldsymbol{\lambda}}^\dagger(\tau)
        \hat{c}_{\nu\boldsymbol{\lambda}}(\tau')
        \hat{c}_{\mu'\boldsymbol{\lambda}}^\dagger(0)
        \hat{c}_{\nu'\boldsymbol{\lambda}}(\tau'') \right\rangle \\ &=
        -\iint d\tau' d\tau'' \mathrm{tr}\big[ \hat{G}_{\boldsymbol{\lambda}}(\tau') \hat{\Gamma}_j^{\perp}(\boldsymbol{\lambda},-\tau'') \\ &\phantom{=} \times \hat{G}_{\boldsymbol{\lambda}}(\tau''-\tau)\hat{\gamma}_i^\perp(\boldsymbol{\lambda},\tau-\tau';\tau',\tau''-\tau) \big].
    \end{split}
\end{align}
The newly introduced vertex $\hat{\gamma}_i^\perp(\boldsymbol{\lambda},\tau_1;\tau_2,\tau_3)$ comes from the diagrammatic expansion (Figure \ref{fig:bubble-diagram-expansion}), which also includes the irreducible crossed-line bubble interaction diagram $\hat{\hat{\Lambda}}_{\boldsymbol{\lambda}}(\tau_1,\tau_2)$ \cite{manybody-bruus-flensberg:2004} 
\begin{align} \label{eq:3p-vertex-dyson}
    \begin{split}
        \hat{\gamma}_i^\perp(\boldsymbol{\lambda},\tau_1;\tau_2,\tau_3) &= \hat{\Gamma}_i^{\perp}(\boldsymbol{\lambda},\tau_1) \\ &\phantom{=} + \mathrm{tr}\bigg[ \hat{G}_{\boldsymbol{\lambda}}(\tau_2) \hat{\hat{\Lambda}}_{\boldsymbol{\lambda}}(\tau_2,\tau_3) \hat{G}_{\boldsymbol{\lambda}}(\tau_3) \\&\phantom{=+} \hspace{45pt} \times \hat{\gamma}_i^\perp(\boldsymbol{\lambda},\tau_1;\tau_2,\tau_3) \bigg].
    \end{split}
\end{align}
Equation (\ref{eq:3p-vertex-dyson}) can be solved self-consistently with appropriate approximations for the irreducible cross-line diagrams \cite{manybody-bruus-flensberg:2004}. Thus, the generalized QGT is obtained by self-consistently solving the interaction vertex (Equation \ref{eq:bethe-salpeter}), which allows solving the vertex correlator $\chi_{ij}(\boldsymbol{\lambda},\tau)$ (Equation \ref{eq:vertex-correl-expand}). From the latter we obtain $\Re\left[ \sigma_{ij}(\boldsymbol{\lambda},\omega) \right]$ by analytical continuation (Equation \ref{eq;anal-cont-mats}) and $\Im\left[ \sigma_{ij}(\boldsymbol{\lambda},\omega) \right]$ from Kramers-Kroing relations (Equation \ref{eq:kramerskronig}). The spectral tensor $T_{ij}(\boldsymbol{\lambda},\omega)$ is then obtained from Equation \ref{eq:Tij-sigma}, from which the proper quantum geometric tensor $T_{ij}(\boldsymbol{\lambda})$ is determined by integrating over frequencies (Equation \ref{eq:spec-to-proper}).

In general, different sub-manifolds can coexist and can be generically coupled, which mathematically is described by their product 
\begin{equation}
    \mathcal{M}
    =
    \prod_{\mu=1}^N\mathcal{M}^{(\mu)},
\end{equation}
%
%where each element $\mathcal{M}^{(\mu)}$ is parametrized by a set of generalized coordinates $\lambda^{(\mu)}_{i_\mu}$, which relate to crystal momentum, or deformation channels related to strain, collective or structural modes. These coordinates label the many-body ground state functions determined for each parameter value. Thus, the distance on this global manifold has sum terms depending on parameters of different sub-manifolds
%
The distance on this global manifold has sum terms depending on parameters of different sub-manifolds
\begin{equation}
    ds^2 = T^{(\mu,\nu)}_{i_\mu j_\nu} d\lambda^{i_\mu} d\lambda^{j_\nu},\\
\end{equation}
where $T^{(\mu,\nu)}_{i_\mu j_\nu}$ are the building blocks for the global QGT
\begin{align}
    \mathcal{Q} &= \begin{pmatrix}
        T^{(1,1)} & \cdots & T^{(1,N)} \\
        \vdots & \ddots & \vdots \\
        T^{(N,1)} & \cdots & T^{(N,N)}
    \end{pmatrix}.
\end{align}
The diagonal blocks $T^{(\mu,\mu)}$ describe the quantum geometry within a
given deformation channel $\mathcal{M}^{(\mu)}$, whereas the off-diagonal blocks $T^{(\mu,\nu)}$ ($\mu\neq\nu$) describe the case where different channels are coupled. Within each block, the tensor structure is determined from the corresponding dressed-vertex responses,
\begin{equation}
    \chi_{i_\mu j_\nu}^{(\mu,\nu)}(\omega)
    \sim
    \left\langle
    \hat{\Gamma}_{i_\mu}^{\perp};
    \hat{\Gamma}_{j_\nu}^{\perp}
    \right\rangle_{\omega}.
\end{equation}
Trivially, if this correlation vanishes for a block $\mu \neq \nu$ these two manifolds are decoupled.

Let us specify a general way to obtain the vertices associated with different many-body fields by introducing an external field source $J_\alpha(\mathbf{q})$ conjugate to a fermionic operator $\hat O_\alpha(-\mathbf q)$, where $\alpha$ is a component within a given interacting channel associated to e.g., spin, charge, or orbital fields. In this case, we can parameterize the Hamiltonian so that $\hat{\mathcal H}[\vec J] = \hat{\mathcal H}_0 - \sum_{\mathbf q,\alpha} J_\alpha(\mathbf q)\, \hat O_\alpha(-\mathbf q)$. Here, $\hat{\mathcal H}_0\equiv\hat{\mathcal H}[\vec J=0]$, while $J_\alpha(\mathbf q)$ is the external field conjugate to the operator $\hat O_\alpha(-\mathbf q)$. The QGT is then determined by the geometry generated by the mapping $\vec{J}\mapsto|\psi [\vec{J}]\rangle$ (or $\vec{J}\mapsto\hat{\rho} [\vec{J}]$ at finite temperatures). In this case, the generalized vertex can be defined as a functional derivative

%Its symmetric part corresponds to the fidelity-susceptibility matrix associated with infinitesimal variations of the external sources as described in Refs. \cite{provost1980riemannian, you2007fidelity, campos2007quantum}. In this case, the generalized vertex can be defined as a functional derivative,

%In the generalized QGT approach, the source amplitudes $J_\alpha(\mathbf{q})$ constitute the coordinates on a parameter manifold that label the many-body ground state wavefunctions $|\psi [\vec{J}]\rangle$ (or density matrices $\hat{\rho} [\vec{J}]$ at finite temperatures), i.e., each configuration of $\vec{J}$ corresponds to a many-body ground state $|\psi [\vec{J}]\rangle$ or density matrix $\hat{\rho} [\vec{J}]$. 

  %, and each source configuration defines a many-body ground state $|\Psi_0[J]\rangle$. The corresponding QGT, evaluated locally at $J=0$, quantifies the infinitesimal variation of these states under changes of the selected collective source. The physical generalized-vertex associated with the source is

%Though not essential to the present formulation, a general motivation for introducing the source $J_\alpha(\mathbf{q})$ comes from Hubbard–Stratonovich transformations, which reveal the physical interaction channels (spins, orbitals, charge) through an auxiliary bosonic field coupled to $\hat O_\alpha(-\mathbf q)$.

\begin{equation*} \label{eq:int_vertex}
    \begin{split}
    \hat{\Gamma}_\alpha[\vec{J}] &= -\dfrac{\delta\hat{G}^{-1}[\vec{J}]}{\delta J_\alpha(-\vec{q})} \\ 
    %&= 2\omega_\alpha J_\alpha(\vec{q})\hat{\mathbb{I}}-
    %\hat O_\alpha(\mathbf q),
    \end{split}
\end{equation*}
where the fermionic momentum-frequency variables of the Green function and the bosonic momentum-frequency transfer $(\vec{q},i\nu_m)$ are left implicit.

%which is used to compute the associated QGT, $T_{\alpha\beta}[\vec{J}]$, from Equations \ref{eq:Tij-sigma} and \ref{eq;anal-cont-mats}. The fermionic momentum-frequency variables of the Green function and the bosonic momentum-frequency transfer $(\vec{q},i\nu_m)$ carried by the source are left implicit in Equation \ref{eq:int_vertex}.
%taking into account that the density operator is represented using the identity matrix $\hat{\mathbb{I}}$.

Let us exemplify the present formulation for the specific case of spin fluctuations. Previously, we have found that spin fluctuations generate a genuine many-body geometric curvature that cannot be otherwise described by bare-band geometric contributions \cite{berryjt-minarro:2026}. For that, we used a formulation based on the geometric tensor defined in momentum space dressed by collective fluctuations. However, the same problem can be treated under the formalism of a generalized QGT. Here, we consider sources $\vec{J}_{\alpha,s}$ that couple the spin bosonic fields to the fermionic spin-density operators given by $\hat O_{\alpha,s}(\mathbf q) = \frac{1}{2} \sum_{\mathbf k} \hat{\vec{c}}^{\dagger}_{\mathbf k+\mathbf q}\,\hat{\sigma}_{\alpha}\, \hat{\vec{c}}_{\mathbf k}$, where $\sigma^{\alpha}$ are Pauli matrices in spin space with $\alpha=x,y,z$. The associated fluctuation-driven curvature can then be reinterpreted in the formalism developed here as the antisymmetric part of the correlations between the generalized interactive vertices defined as $\hat{\Gamma}_\alpha[\vec{J_{\alpha,s}}](\imath\omega_n)$.

We can similarly treat the Jahn-Teller vibronic problem under the light of the generalized QGT approach \cite{khomskii2014,bersuker2013jahn,jahn-tel:1937,tsukerblat2006}, where electronic orbitals couple with lattice distortions. The latter are parametrized by a vector \(\mathbf{Q}\), representing the coordinates of the ionic displacements in the space of normal modes, whose components correspond to allowed irreducible representations \cite{bsk-pol:1989,bersuker2013jahn}. For example, if the relevant distortions correspond to $E_g$ representations, as in $E\otimes e$ and $T \otimes e$ vibronic couplings \cite{bsk-pol:1989,minarro2022prb}, then $\vec{Q} = (Q_\theta,Q_\phi)$ where $Q_\theta$ is the tetragonal distortion axis and $Q_\phi$ is the orthorhombical distortion axis. One can then directly regard \(\mathbf{Q}\) as a physical nuclear coordinate manifold and define manybody ground state functions parametrized by $|\Psi_\mathbf{Q}\rangle$, making \(\mathbf{Q}\) the natural geometric coordinates.

These distortions can be described by the following Hamiltonian \cite{streltsov2020jahn,jtpolaron-takada:2001,jtpolaron-takada:2007,jtpol-trugman:2004,kugel1982jahn}
\begin{align}
    \hat{\hamil}_\vec{Q}^{jt} &= \dfrac{g_\eta}{\sqrt{3}}\hat{\vec{c}}_\vec{Q}^\dagger \hat{\lambda}^\eta \hat{\vec{c}}_\vec{Q} Q_\eta + \omega_\eta Q^\eta Q_\eta \hat{n}_\vec{Q},
\end{align}
where $\eta$ represents the distortion component, $g_\eta$ is the coupling constant, $\omega_\eta$ is the natural frequency of the vibronic mode and $\hat{\lambda}^\eta=\lim_{\vec{Q}\to 0}\partial_\eta \hat{V}(\vec{Q})$ being $\hat{V}$ the anion-electron electrostatic potential \cite{bsk-pol:1989}. In some cases, e.g. in dynamic Jahn-Teller regimes \cite{iwahara2024dynamic}, these distortions are described by quantum vibronic models involving $\hat{Q}_\eta = \hat{b}_\eta + \hat{b}_\eta^\dagger$, thus requiring quantum operators living in the phonon Hilbert space \cite{jtpolaron-takada:2001,jtpolaron-takada:2007,minarro2025emergent}. Here, though, we consider a semi-classical representation, where the parametric vertex is determined from a semi-classical representation of Jahn-Teller coordinates
\begin{align}
    \hat{\Gamma}_\eta(\vec{Q},\tau) &= \dfrac{g_\eta}{\sqrt{3}} \hat{\lambda}_\eta + \partial_\eta\hat{\Sigma}_\vec{Q}(\tau) + 2\omega_\eta Q_\eta \hat{\mathbb{I}},
\end{align}
%
%where the last term is an identity matrix, $\hat{\mathbb{I}}$. In this case, the vertex correlator is defined through an effective vertex given by $\chi_{\eta\xi}(\vec{Q},\tau) = -\left\langle  \hat{\tilde{\Gamma}}_\eta(\vec{Q},\tau)  \hat{\tilde{\Gamma}}_\xi(\vec{Q},0) \right\rangle$, where $\hat{\tilde{\Gamma}}_\eta(\vec{Q},\tau) = \dfrac{g_\eta}{\sqrt{3}} \hat{\lambda}_\eta + \partial_\eta\hat{\Sigma}_\vec{Q}(\tau)$.

If we restrict to the particular case where the self-energy is independent of distortions $\hat{\Sigma}_\vec{Q}(\tau) = \hat{\Sigma}(\tau)$, then $g_\eta = g_\xi \equiv g_{jt}$ and $\omega_\eta=\omega_\xi \equiv \omega_0$, where the vibronic coupling strength is the same for all Jahn-Teller modes corresponding to same irreducible representation \cite{bersuker2006jt}. In the mean field approximation, the perpendicular vertex correlator is reduced to
\begin{align*}
    \chi_{\eta\xi}(\tau) &= -\dfrac{g_{jt}^2}{3}\mathrm{tr}\left[ \hat{\lambda}_\eta \hat{G}_\vec{Q}(\tau)\hat{\lambda}_\xi\hat{G}_\vec{Q}(-\tau) \right]
\end{align*}
In this situation, the interactive vertices inherit the time-reversal symmetry of the electrostatic Jahn-Teller potential and, therefore, the antisymmetric response vanishes. This leads to a vanishing curvature, while the metric is Euclidean if we assume a single vibronic channel (e.g. $T\otimes e$), or non-Euclidean if we consider both $T\otimes e$ and $T\otimes t$ problems \cite{streltsov2022interplay}. Note that the vanishing curvature is local, which distinguishes from the appearance of $\pi$-Berry phases, which originate from global holonomies due to the nuclear motion around conical intersections \cite{bersuker2006jt, khomskii2020orbital, khomskii2014}.

%This condition may apply to dynamical Jahn-Teller effects, in which quantum fluctuations restore the symmetry of the ground state \cite{iwahara2024dynamic}.

%\section{Conclusions}

%The dynamic properties of quantum particles along any arbitrary parameter space can be modeled using differential geometry concepts. This dynamics is ruled by geometric tensors, metric and curvature, the real and imaginary parts of the quantum geometric tensor, the one that relates the gauge independent infinitesimal distance of wave-functions and infinitesimal variations of parameters. The computation of these tensors require the Berry connections. These objects needs the knowledge of wave-functions of every band to be computed. On strong-correlated systems this computation is extremely huge because of particle-particle interactions. However, on this work, it has been demonstrated other way to arrive to these tensors.

In view of these considerations, the generalized formulation of quantum geometry presented here provides a diagrammatic route to incorporate many-body vertex corrections into quantum geometry, thereby complementing the growing efforts to extend quantum-geometric concepts to interacting systems \cite{yu2025quantum,jin2026experimental, guan2026exploring, chen-dressedberry:2022}. At the same time, our formulation places electromagnetic responses, collective electronic channels, and structural distortions within a common many-body response framework.\\

\textbf{Acknowledgments} This work was supported by Projects No. PID2023-152225NB-I00 and Severo Ochoa MATRANS42 (No. CEX2023-001263-S) of the Spanish Ministry of Science and Innovation (Grant No. MICIU/AEI/10.13039/501100011033 and FEDER, EU). We also acknowledge fruitful discussions with Prof. Min-Fong Yang from Tunghai University, Taiwan.

\appendix

\onecolumngrid

\twocolumngrid

\bibliography{bibliography} % Entries are in the refs.bib file
\bibliographystyle{apsrev4-2} % We choose the "plain" reference style

\end{document}